\documentstyle[11pt,psfig,aaspp4,flushrt]{article}
\def\be{\begin{equation}}
\def\ee{\end{equation}}
\def\etal{{\it et al. }}

\def\micron{$\mu$m}
\slugcomment{Version 4.0, Feb 2001}
\begin{document}

\title{The Optical/Infrared Astronomical Quality of High Atacama Sites.
I. Preliminary Results of Optical Seeing}
\author {\it Riccardo Giovanelli$^1$, Jeremy Darling$^1$, Marc Sarazin$^2$,
Jennifer Yu$^4$, Paul Harvey$^3$, Charles Henderson$^1$, William Hoffman$^1$, 
Luke Keller$^1$, Don Barry$^1$,
James Cordes$^1$, Stephen Eikenberry$^1$, George Gull$^1$, Joseph Harrington$^1$,
J. D. Smith$^1$, Gordon Stacey$^1$, Mark Swain$^1$}
\affil{$^1$Department of Astronomy, Cornell University, Ithaca, NY 14853}
\affil{$^2$European Southern Observatory, Garching bei M\" unchen, 
D--85748 Germany}
\affil{$^3$Department of Astronomy, University of Texas, Austin, TX 78712}
\affil{$^4$Department of Earth and Atmospheric Sciences, Cornell University, Ithaca, NY 14853}

\hsize 6.5 truein

\begin {abstract}
The region surrounding the Llano de Chajnantor, a high altitude plateau in the 
Atacama Desert in northern Chile, has caught the attention of the astronomical 
community for its potential as an observatory site. Combining high elevation and extremely low atmospheric water content, the Llano has been chosen as the future 
site of the Atacama Large Millimeter Array. We have initiated a campaign to 
investigate the astronomical potential of the region in the optical/infrared. 
Here, we report on an aspect of our campaign aimed at establishing 
a seeing benchmark to be used as a reference for future activities in the region.
After a brief description of the region and its climate, we describe the 
results of an astronomical seeing campaign, carried out with a Differential Image 
Motion Monitor that operates at 0.5 \micron ~wavelength. The seeing at the Llano 
level of 5000 m, measured over 7 nights in May 1998, yielded a median FWHM of 
1.1". However, the seeing decreased to 0.7" at a modest 100 m gain above the plateau (Cerro Chico), as measured over 38 nights spread between July 
1998 and October 2000. Neither of these represents the best seeing expected in
the region; the set of measurements provides a reference base for
simultaneous dual runs at Cerro Chico and at other sites of interest
in the region, currently underway. A comparison between simultaneous 
measurements at Cerro Chico and Cerro Paranal indicates that seeing at Cerro Chico 
is about 12\% better than at Paranal. The percentage of optically photometric
nights in the Chajnantor region is about 60\%, while that of nights useful for
astronomical work is near 80\%.
\end{abstract}

\noindent {\bf Subject Headings}: Astronomical instrumentation, methods and techniques:
atmospheric effects, site testing.

\section {Introduction}

Outer space provides the ideal setting for astronomical
instruments. Freeing them from the limitations imposed by
our planet's atmosphere, observations of high angular resolution and
in spectral bands to which the atmosphere is opaque have become possible.
However, because of the steep costs of hauling large masses to space, very 
large collecting area telescopes have remained the domain of ground--based
observatories. In addition to high sensitivity, large apertures, aided by 
adaptive optics techniques, allow high angular resolution images to be 
obtained from the ground, which are most effective in sites with good 
astronomical seeing.
Especially in the infrared, atmospheric opacity and emissivity can be 
minimized by placing observatories at high altitude or in extremely cold 
locations. The South Pole has witnessed strong growth in the number and 
quality of astronomical installations in recent years. Similarly, high and 
dry sites have become increasingly attractive, in spite of the
operational difficulties that arise with high altitude.

The Atacama Desert in northern Chile is one of the driest regions on Earth. It 
lies between the Coastal Cordillera to the west and the Andes to the east. The 
region of the Altiplano to the east of the Salar de Atacama known as Llano de 
Chajnantor, a plateau of altitude near 5000 m, was selected by the U.S. National 
Radio Astronomy Observatory (NRAO) as the future site for its Millimeter Array 
Project, while a neighboring plateau, {\it Pampa La Bola}, was selected by the 
Nobeyama Radio Observatory of Japan for its Large Millimeter and Submillimeter 
Array (LMSA) Project. These sites are within a few km from 
each other, at elevations between 4800 and 5050 m above mean sea level. 
Successively, the Millimeter Array Project has evolved into a U.S.--Europe 
consortium to build the Atacama Large Millimeter Array (ALMA). Other radio
and optical Astronomy consortia are under development for operation in this 
region, which has the potential for expanding into a major world astronomical
center. Cerro Paranal, the site of the European Southern Observatory's (ESO) Very
Large Telescope Project, is about 300 km to the southwest, on a peak on the
Pacific coastal range (Coastal Cordillera).

The climatic qualities that make the Atacama region especially attractive
to astronomers extend over a latitudinal band a few hundred kilometers wide, 
about the Tropic of Capricorn. The access to good quality services and good 
communications further focuses attention on the regions in the vicinity of the 
cities of Antofagasta, Calama and the village of San Pedro de 
Atacama. The presence of the VLT and the likely establishment of major national 
and international research centers, such as ALMA and the LMSA, adds promise of 
scientific and operational synergism to the region. The government of Chile has
legislated the protection of an area which includes the Llano de Chajnantor, 
the Pampa La Bola and the surrounding peaks, as a {\it National Science Preserve}.

Several institutions are carrying out measurements to ascertain the 
characteristics of the region for astronomical observations
in the radio and submillimeter parts of the spectrum. We have initiated a 
campaign to characterize the region for optical and infrared
astronomical observations. In the following two sections, we present a 
brief description of the region and its climate. In  Section 4, the seeing 
measurements and synoptic results are described, followed by brief remarks on 
photometric conditions. We conclude with a discussion on plans and prospects 
for future work. In a companion paper (Giovanelli \etal ~2001; Paper II), 
we present the results of measurements regarding the water vapor content and its 
implications on infrared transparency and background emissivity.

\section {Description of the Region  \label{region}}

\begin{figure}
\plotone{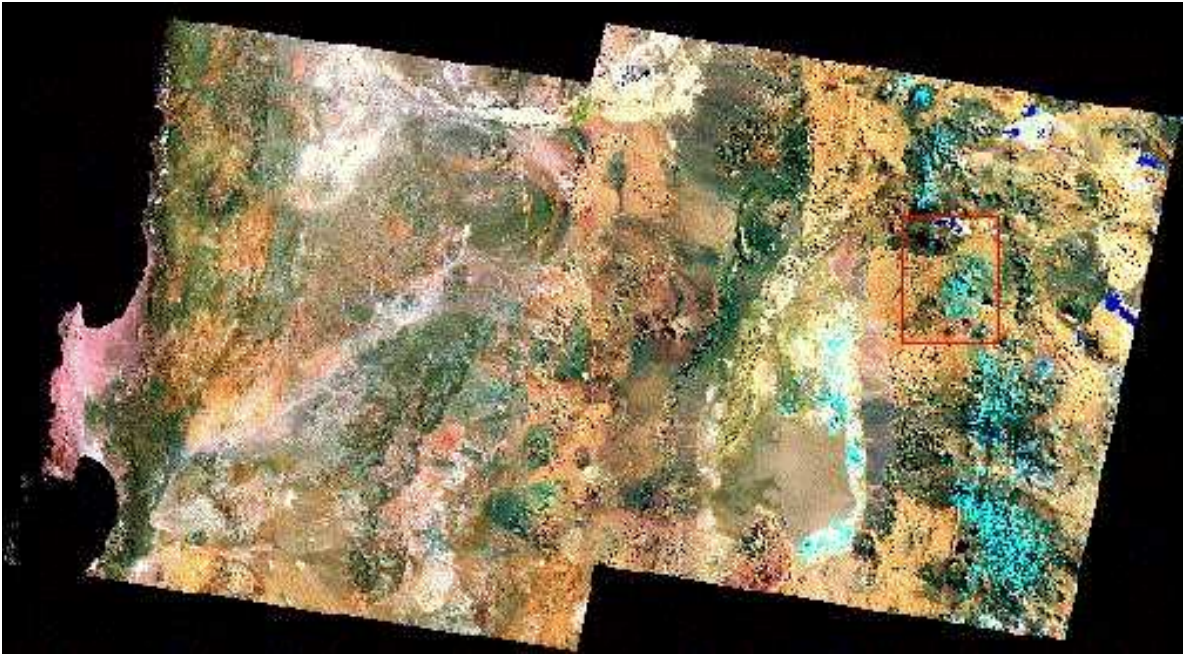}
\caption{Satellite composite view of the Atacama region. The image
encloses a region of approximately 400 by 200 km, extending from
the Pacific coast to the West (left) to the Andean divide to the East. The 
city of Antofagasta is just to the South of the anvil shaped Mejillones 
Peninsula. Calama and Chuquicamata are located slightly North of the 
center of the image, while the Salar de Atacama is the brown region to 
the SE of center, bordered on its eastern side by light blue hues. 
The red square box is centered on the Chajnantor Plateau, better seen 
in Figure \ref{satview}. The NE side of the image includes 
Bolivian territory; Argentina is just off the eastern (right)
edge.
\label{jen1}}
\end{figure}

The Salar de Atacama (23$^\circ$ 30' S, 68$^\circ$ 15' W) is the 
largest salt flat in Chile. A basin with median elevation near 2300 m, 
it extends for more than 100 km from north to south, lies 200 km inland 
from the Pacific coast and is located about 1600 km north of the capital 
city of Santiago. The Salar is bounded to the west by the Domeyko mountain 
range, with peaks as high as 4300 m (Cerro Quimal), and to the 
east by the main Andean magmatic arc. Among the highest Andean peaks in 
the vicinity are Llullaillaco (6723 m) south of the Salar, and Acamarachi 
or Pili (6046 m), while among the most spectacular are Licancabur (5950 m) 
and L\' ascar (5592 m). Figure \ref{jen1}~ is a composite satellite image 
of a region of approximately 400 by 200 km, extending from the Pacific 
coast on the west to the Andes on the east. The image is the 
combination of three frames taken through different broad--band filters: 
0.5 \micron ~(blue), 1.0 \micron ~(green) and 1.5 \micron ~(red). Data 
were acquired on July 20, 1985 by the Landsat 5 imaging satellite. The 
Salar de Atacama is the region filled with muds and evaporitic deposits 
outlined in light brown to the right and below of center, bordered on 
its right by turquoise hues (salt deposits and ponded water). To the east, 
where the volcanoes of the Andean magmatic arc rise above 5000 m, moisture 
has been intercepted by the high peaks and precipitated as snow on the 
plateau (represented by bright blue).  The city of Antofagasta is to the 
south of the anvil--shaped Mejillones Peninsula, at the lower left in the
figure.

Around the Salar, human settlements are only found in the vicinity of oases and 
they are generally quite small. The largest among them includes the village of 
San Pedro and its environs, with a combined population of about 1,200. San Pedro 
is located at the northern end of the Salar de Atacama, just to the west (left) of 
the red outline in Figure \ref{jen1}. No humans reside between San Pedro and the 
Bolivian border, 50 km 
to the northeast, nor between San Pedro and the Argentine border, 160 km to 
the east. The nearest large urban center is Calama--Chuquicamata, with a 
combined population of about 150,000, 100 km to the northwest of San 
Pedro, beyond the Domeyko range. Calama--Chuquicamata is at the center of 
the most important copper mining zone in the South American continent (just
south of the upper boundary, center, of the image in Figure \ref{jen1}). Its 
airport is connected by several daily flights to other Chilean cities, such as 
Antofagasta, Santiago, Copiap\' o and La Serena. Antofagasta, the regional capital 
with a population of one--third of a million, is on the coast 200 km to the 
west of Calama. The nearest town with population exceeding 1000 to the east 
of San Pedro is across the Andes, some 500 km away. To the north, 
the nearest settlements are at Ollag\" ue, at 300 km, and Uyuni, 400 
km away; neither exceeds the size of San Pedro. To the south, only 
a few villages much smaller than San Pedro dot the edges of the Salar. 
From an urban viewpoint, Atacama is thus a region of extreme isolation.

Traffic is however increasing, both because the colonial and archaeological 
patrimony and natural beauty of the surroundings make San Pedro a favorite 
tourist destination, and because the international Andean road through the Paso 
de Jama to Argentina is now paved. This road will likely become one of the 
main arteries between the continental heartland (northern Argentina, Paraguay and 
southern Brazil) and the Pacific Ocean. The use of urban lights will undoubtedly 
increase in San Pedro. However, the likelihood of tumultuous future development 
of San Pedro or of any of the neighboring villages into major desert centers, as 
has occurred in the American Southwest, is remote. Sizable sources of water in 
the region are absent. In spite of the high altitude of the Andean peaks, there 
are no extensive glaciers in this region, and precipitation is extremely rare. 
The villages 
in the Atacama Desert are unlikely to pose a serious light pollution threat in 
the foreseeable future.

The quality of services in the region is high, a result of the conspicuous 
presence of major mining interests. Roads are good, supply,
communications and financial services are excellent, medical services abound 
and unique expertise in the special circumstances of high altitude physiology 
exists. For example, delivery of supplies, from water to cryogens, to remote, 
high altitude sites is a routine practice; engineering contracting companies 
expert in planning and construction of facilities at high altitude are numerous; 
and international meetings on high altitude medicine are held in the region (e.g. 
Iquique, November 2000).

\begin{figure}[t]
\plotone{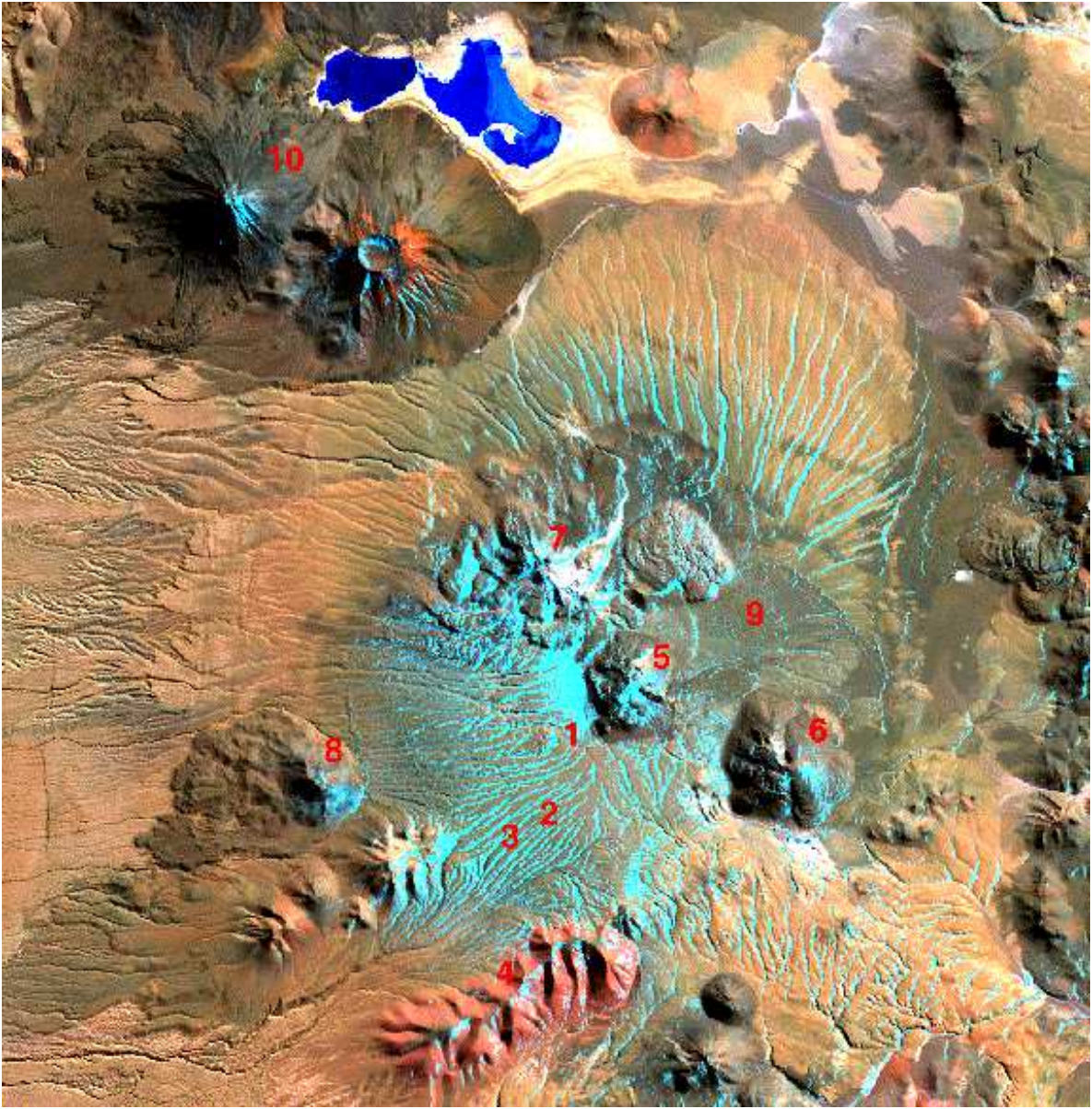}
\caption{Satellite composite view of the Chajnantor Plateau region. 
The image subtends an area of approximately 1500 square km.
North is up and left is West. It corresponds roughly to the
region within the red outline in Figure \ref{jen1}.
Labeled features are identified in Section 2.\label{satview}}
\end{figure}

The Llano de Chajnantor is located about 40 km to the east of San Pedro, 
and nearly 3000 m higher in elevation. The region is illustrated in Figure 
\ref{satview}, an area of 1560 km$^2$ roughly corresponding to that
within the box outline in Figure \ref{jen1}. The Salar de Atacama lies 
to the southwest of the region displayed in Figure \ref{satview}~and the 
town of San Pedro is at the northern end of the Salar, about one half 
map-width off the western margin of the map. Several notable features 
are labeled. The two volcanoes {\it Licancabur} (5950 m) and {\it Juriques} 
(5750 m, with the red--rimmed crater) are in the northwest. Licancabur 
(label 10), is a prominent, symmetric cone and marks the southwestern vertex 
of the Chile--Bolivia political border. To the northeast of the two volcanoes 
are the two lagoons jointly known as {\it Laguna Verde}, in Bolivian territory. 
Near the center is {\it Cerro Chajnantor} (label 5, 5700 m); to the southeast 
of it is {\it Cerro Chasc\' on} 
(label 6, 5750 m) and to the northeast is {\it Cerro Toco} (label 7, 5650 m). 
A sulphur--rich ground surface, east of Toco's summit, is responsible
for the white coloring in that part of the image. Light blue implies the 
presence of snow. The red--colored ridge running northeast to southwest 
(label 4), is known as {\it Cerros de Honar} (5400 m), and the mountain 
labeled 8 is {\it Cerro Negro} (5100 m). The chain of hills separating 
Negro and Honar are known as {\it Cerros de Mac\' on} (5000 m). The Llano 
de Chajnantor is the area between Chajnantor, Chasc\' on, Honar and Negro. 
Its average elevation is 5000 m. Both NRAO and ESO maintain ALMA site 
testing equipment in the plateau, at the location identified 
by the label 2 in the map. The region north of Chasc\' on and 
east of Chajnantor is known as {\it Pampa La Bola}. At an average 
elevation of 4800 m, it hosts testing equipment of the Nobeyama Radio 
Observatory for the LMSA project, of Japan, near the label `9'. Finally, 
label `1' identifies {\it Cerro Chico} (5150 m), 
the site at which most of the seeing tests described in this 
report were conducted; and label `3' identifies the location of the 
radiosonde launch facility, jointly operated by Cornell, NRAO, ESO, 
the Harvard--Smithsonian Astrophysical Observatory (SAO) and the Nobeyama
Radio Observatory (since 2000), which produced 
much of the atmospheric water vapor data  discussed in Paper II.

For scale, the distance between the summits of Juriques and Chajnantor 
is 17.5 km. The study area is dominated by volcanic surfaces that range in
age from approximately $~10^6$ to $10^7$ years. These features are remarkably 
well-preserved due to the aridity of the region. The Llano de Chajnantor is 
part of a circular area about 12 km in diameter centered on Cerro Chajnantor, 
which is a Quaternary andesitic dome.  The surrounding surfaces include 
Plio-Quaternary andesites and Quaternary ignimbrites. The drainages lead 
radially outward from Cerro Chajnantor. The topography rapidly loses elevation 
moving 30 km west from the edge of the Llano at 5000 m to the Salar at 2300 m.  
Among the peaks in the area, Honar rises 400 m above the plateau, Chico rises 
less than 150 m, Chajnantor and Toco both rise about 650 m, and Chasc\' on is 
the highest local point at 800 m above the Llano. Images of the region can be
viewed at {\it http://www.astro.cornell.edu/atacama}.

The northern Chilean Andes include a high concentration of geologically active 
volcanoes. In the vicinity of Chajnantor, the only conspicuous and current 
activity takes place at Volc\' an L\' ascar,  located about 30 km south of Honar. 
L\' ascar's last mayor eruption took place in 1993, producing a tephra cloud 
which the winds carried principally eastward. It produced volcanic ash deposits 
3 cm thick 50 km downwind from its crater (Gonz\' alez--Ferr\' an 1994). A
minor eruption of the same volcano took place on July 20, 2000, during one
of our campaign runs. The 
likelihood of other volcanic eruptions in the region are discussed by Gardeweg 
(1996), while a thorough survey of the volcanology in Chile is 
given by Gonz\' alez--Ferr\' an (1994).

\section {Remarks on Climate}  

\subsection {Conditions at the Llano de Chajnantor}
The work 
of Dieter Schmidt (1996) provides the most recent and complete compendium 
of climatological conditions of the Atacama. See also Cook (2000) and 
Fuenzalida (1984).

The large--scale climatological patterns of the region are associated
with the persistence of the Southeast Pacific subtropical anticyclone. 
An important driving force of the regional weather pattern
is the extremely high fraction of the solar radiation flux
reaching the ground, due to the tropical location and the high atmospheric 
transparency. The heat input by radiation is released through convective 
and advective processes, due to the lack of humidity to form latent heat, 
with the wind playing an important role in heat transport. Low altitude 
winds in this region are generally driven by the temperature differences 
between the Pacific coast and the Andean region, exhibiting a clear 
diurnal cycle. Both the diurnal and seasonal cycles of the wind speed at 
the Llano de Chajnantor, as recorded at the ALMA site between 1995 and 
2000, can be viewed in the composite figures of the continuously updated 
NRAO web site  {\it http://www.tuc.nrao.edu/mma/sites/Chajnantor/data.c.html} 
(Radford 2000). 

Between April 1995 and October 2000, the 25\%, 50\% and 75\% wind speed 
quartiles were respectively 3.1, 6.4 and 10.3 m s$^{-1}$ at the plateau level. 
Higher speeds occur during daytime hours, rising rapidly a few hours after 
sunrise in mid--winter and in the mid--afternoon during Summer. Sustained 
speeds can exceed 15 m s$^{-1}$. The nighttime median wind speed at 4 meters 
off the ground is about 4 m s$^{-1}$, a relatively
benign condition which is, in fact, lower than that observed at Mauna
Kea. In both Summer and Winter, wind speed drops precipitously about 
one or two hours after sunset. A seasonal pattern is also clear, with wind
speeds decreasing in the austral Summer months. Wind direction is also tied
to the seasonal cycle: the flow is nearly always from the west between
April and early December, while the wind direction becomes more variable in 
the Summer months, when conditions locally referred to as ``Bolivian Winter'' 
often ensue. These are associated with the inflow of moist Amazonic air from 
the NE. Hence the local name. Wind storms lasting 1--2 days with wind speeds
exceeding 25 m s$^{-1}$ occur a few times per year, especially during
winter months. These events are well correlated with high wind speeds at the
200 mb level, the occurrence of which can be anticipated several days in
advance with good reliability. In that respect, the forecasts posted by M. 
Sarazin in 
{\it http://www.eso.org/gen-fac/pubs/astclim/forecast/meteo/ECMWF/long/chajnantor}
have proved extremely valuable for our campaign activities.

High winds can be of hindrance to astronomical operation not only because
of their mechanical effects, but also because they can increase the density
of dust found in suspension in the atmosphere. Near valley floors, in
fact, dust storms are frequent, due to the sandy characteristics of the 
soil. In the Chajnantor region, however, the ground is generally a mixture 
of rock and gravel, and little dust is raised even by the strong daytime
winds. The valley dust layer is seldom seen to rise above the 4500 m
level. Quantitative determinations of atmospheric dust content at
Chajnantor are planned.

The ground temperature distribution at the Llano de Chajnantor, as recorded 
at the ALMA site testing station, can also be appreciated in Radford (2000)
both in its seasonal and diurnal characteristics. The historical 25\%, 50\% 
and 75\% quartile values for ground level temperature at the ALMA monitoring 
site are respectively 2.9 C, -2.6 C and -7.8 C. The diurnal cycle has an 
amplitude of 13--14 C, while the seasonal cycle has a slightly 
lesser amplitude of 10--12 C. The temperatures at the plateau 
are not extreme, reaching as high as 15 C during summer days and 
seldom falling below -15 C during winter nights. Night temperatures
vary little: between 0 and 10 hours UT (approximately 7:30 p.m. and 5:30 a.m.,
local time), the average temperature drop is 2.5 C, i.e. approximately 0.25 C
hr$^{-1}$.

At San Pedro de Atacama, precipitation is extremely low, typically 
50 mm or less per year. At the plateau it precipitates somewhat more 
frequently, generally in the form of snow and more likely when Bolivian 
Winter conditions occur. Snow accumulation seldom exceeds a few cm, 
although wind--driven drifts may accumulate layers of substantial thickness
in sheltered depressions of the terrain. In shaded areas, melting and
sublimation is slow and relatively small amounts of snow can last 
weeks or months.  

{\it El Ni\~ no--La Ni\~ na} events, of 12 to 24 months duration, produce 
very significant alterations, superimposed on the annual cycle of atmospheric 
conditions. Severe {\it El Ni\~ no} events recur about once per decade, 
based on records of the last two centuries, while more moderate events 
take place about twice as frequently (Rodbell \etal 1999; see Neelin \& Latif
1998 for a recent description of El Ni\~ no dynamics). The last  
{\it El Ni\~ no--La Ni\~ na} event took place between December 1997 and 
late 1998. It is reported to have been among the most severe since the 
middle of the XIX century (McPhaden 1999). 
During this event, the sea surface temperature anomaly in equatorial waters 
approached 5 C, while the average event produces deviations 
of 1.5 C or less. The effects of the 1997--98 episode resulted in  
higher humidity, increased atmospheric opacity and worse seeing conditions
than normal, as we discuss later.

\subsection {The Mean Atmosphere Above the Llano de Chajnantor} 

Since sites above the plateau will provide the best conditions for optical/IR 
telescopes, it is important to know what the differential
meteorological conditions are between the relatively well sampled plateau
level and the higher elevations.
Continuous monitoring of the meteorological parameters at elevations above 
the Llano de Chajnantor has only started in October 2000, thus no reliable
record exists of the differential conditions among sites. Some indications 
of the conditions that might be expected at higher elevations can however 
be garnered from radiosonde launches that have taken place since 1998, as 
we report elsewhere in greater detail (Paper II). In this Section, we restrict 
our presentation to median atmospheric profiles obtained from 106 radiosonde 
launches. 

We have selected 30 sondes launched in nighttime hours (UT 01 to 11 hours;
local midnight takes place at 04$^h$ 31$^m$) and 65 sondes launched in daytime
hours (UT 12 to 21 hrs). The sondes were launched between the months of
April and early December, although the month of November is overrepresented 
(14 of 30 night launches and 25 of 65 day launches); thus they are representative 
of Winter and Spring, which is the period when the best astronomical observing 
conditions are found in the region.

The median atmospheric profiles these data yield are very instructional.
Profiles are averaged over intervals of 10 m in altitude and shown in Figures 
\ref {mean_temp}~ and \ref{mean_wind},  It should be kept in mind that these 
profiles refer to the free atmosphere and, due to prevailing winds, generally 
the sonde flights sample an inclined path that takes them to the east of the 
plateau. Conditions may thus be slightly different than over the Llano de 
Chajnantor proper. In addition, conditions at the peaks of mountains may be 
slightly different from those of the free atmosphere, as they will be affected 
by local topography. Nonetheless, these mean profiles give us at least a rough 
idea of the conditions we may encounter at various sites above the plateau.

Figure \ref {mean_temp}~ indicates that at night there is a conspicuous inversion
layer near the ground, which is at an elevation of 5000 m. The air temperature 
increases by several degrees up to about 100 m above the ground. This is a 
consequence of the rapid radiative cooling of the ground after sunset. Between 
5100 and 6000 m, the median temperature profile is ragged, due to the frequent 
occurrence of temperature inversions. Above 6 km elevation, a closer to normal 
tropospheric cooling trend with altitude takes over, at a rate of about 0.7 C 
per 100 m, somewhat lower than the adiabatic lapse rate of 0.98 C per 100 m. 
Conditions at peaks a few hundred meters above the plateau will not be harshly 
different than conditions at the plateau itself. The lapse rate during daytime 
is steeper in the lower 500 m, but median Winter temperatures above -10 C can 
be expected at all accessible ground sites. 
Quartile profiles (dotted lines) in Figure \ref{mean_temp}~ refer to nighttime. 

Figure \ref{mean_wind}~  illustrates the median wind profile. Day and night data 
show, overall, similar profiles from an altitude of 6 km up, the wind speed 
increasing from about 10 m s$^{-1}$ at 6 km to about 25 m s$^{-1}$ at 12 km. 
Differences are most important near the ground, especially between 5 and 6.5 km 
altitude. During day time, the median wind speed at the ground is 7--8 m s$^{-1}$; 
it increases to about 10 m s$^{-1}$ by 5.3 km, and to 12 m s$^{-1}$ by 6.0 km. 
The night time profile is steeper. It rises from about 4 m s$^{-1}$ near the 
ground to about 8  m s$^{-1}$ at altitude 5.3 km, then hovers near 10--11 
m s$^{-1}$ up to 6.2 km. The wind speed at any peak fully exposed to the free 
atmosphere should not be any less than the free atmosphere wind shown in Figure 
\ref{mean_wind}. Thus, night time wind speeds at Cerro Honar (5400 m) should be 
expected near 8 m s$^{-1}$, or twice as high as those at the plateau level, 
while at higher peaks such as Toco, Chajnantor and Chasc\' on, mean speeds of 
the order of 10 m s$^{-1}$ should be expected. Actual wind speeds will depend
critically on the characteristics of the air flow and the local topography.
The profile of the 75\% quartile of night data does however indicate that most 
of the time median wind speeds should not exceed 12--15 m s$^{-1}$. Such wind 
speeds are usually considered to be safely within operational limits at modern 
observatories, although direct testing of gusting conditions is necessary.

\section {Seeing}  

\subsection {Definitions and Terminology} 

Astronomical seeing is the distortion of the radiation wavefront of a cosmic
source, produced by variations in the index of refraction of air. In conditions
that are conducive to astronomical observations, pressure and water vapor
fluctuations are negligible, thus variations in the index of refraction result
primarily from thermal fluctuations associated with turbulent air flow.
The {\it temperature structure function} of an atmospheric layer at altitude
$h$ above the ground is defined in the form of a covariance
\be
D_T(r,h) = < \Delta T(r,h)^2>    \label{eqDT}
\ee
where $\Delta T (r,h)$ is the temperature difference between two points 
separated by a distance r, at constant altitude. Over a wide range of scales
--- between a minimum of a few mm below which turbulence is damped out and 
a maximum size of km, at which turbulence is injected for example by flow 
past an obstacle such as a mountain range --- within the framework of
Kolmogorov's treatment $D_T$ is generally described by a power law of the form
\be
D_T(r) \propto r^\beta      \label{dtprop}
\ee
where $\beta\simeq 2/3$ (Tatarski 1961). The proportionality 
in equation \ref{dtprop} ~is mediated by the scale--independent 
{\it temperature structure parameter} (Roddier 1981):
\be
C_T^2(h) = {D_T(r)\over r^{2/3}}
\ee
which relates to the {\it refractive index structure parameter} via
\be
C_n^2(h) = C_T^2(h) ~\bigl[80\times 10^{-6}~P(h)/T^2(h)\bigr]^2
\ee
where the units of $C_n^2$, pressure $P(h)$ and temperature $T(h)$ are 
respectively m$^{-2/3}$, mb and K.

The spatial coherence scale of atmospheric turbulence is expressed by the
{\it Fried parameter} $r_\circ$ (in m):
\be
r_\circ = \Bigl[ ~0.423~ k^2~\sec(ZA)~\int_{h_\circ}^\infty C_n^2(h)~dh~
\Bigr]^{-3/5} \label{eqrnot}
\ee
where $k$ is the wavenumber in m$^{-1}$, $ZA$ the zenith angle of the line 
of sight and $h_\circ$ the elevation of the telescope above the ground in m.

Finally, the full--width at half--maximum of a stellar image  
at a wavelength $\lambda$ is
\be
\theta_{fwhm} = 0.98 {\lambda \over r_\circ}
\ee
In other words, for a telescope of diameter $D$ turbulence degrades the
image resolution from $\theta_{diff} \simeq \lambda/D$ to the diffraction 
limit which would be observed with a telescope of aperture $r_\circ$,
i.e. $\theta_{fwhm} \simeq \lambda/r_\circ$. Since $r_\circ \propto k^{-6/5}$, as
indicated by Equation ~\ref{eqrnot}, and $\lambda=2\pi/k$, then
\be
\theta_{fwhm} \propto \lambda^{-0.2}  \label{eqtl}
\ee
All layers of the atmosphere contribute to seeing, as illustrated by Equation
~\ref{eqrnot}. 
Turbulence originating in the telescope enclosure is referred to as 
{\it dome seeing}. In our site survey measurements, dome seeing, as well as that
arising from temperature differences between the mirror and the air, are 
null as our telescope has extremely small thermal capacity and it operates 
outdoors. The seeing that originates from convective flows between the ground 
and the few meters above it, and by turbulence arising from air flow between
observatory buildings, is generally referred to as {\it ground surface
layer seeing}. Seeing that originates from large scale interactions
between the ground and the lower atmosphere, such as air flowing over 
topographic irregularities, is referred to as {\it boundary layer seeing}.
Hereafter, we will often refer to the joint effect of ground surface layer 
and boundary layer turbulence as boundary layer seeing. The thickness of 
the boundary layer varies between 200 and 2000 m, depending on the local 
topography, latitude, wind velocity, amplitude of the diurnal thermal cycle 
and other environmental parameters. Katabatic flows --- winds associated with 
the downhill flow of colder air from mountain tops --- can be considered a 
subset of the boundary layer component. They primarily affect locations on 
the slopes and near the foot of substantial peaks. Finally, the contribution
to seeing of that part of the atmosphere which is above the boundary layer, 
in which air flow is generally almost laminar, will be referred to as
{\it free atmosphere seeing}.

\subsection {Differential Image Motion and Seeing}

Our seeing measurements are carried out with a device known as {\it 
Differential Image Motion Monitor} (DIMM). It measures wavefront
slope differences between two pupils on a common telescope mount, as described 
by Sarazin \& Roddier (1990). We have deployed two DIMMs in Atacama. The first
was obtained on loan from the European Southern Observatory (hereafter DIMM--4,
as referred to at ESO), thanks to the kind
disposition of Marc Sarazin and Riccardo Giacconi. The second was assembled
at Cornell University (hereafter DIMM--C1). They
each consist of of an 11--inch Celestron (C11) telescope, an SBIG ST5 CCD 
camera, control equipment, an aperture mask and a laptop computer equipped 
with specialized software written by Peter Wood and ported to ST5 by Marc 
Sarazin. The aperture
mask is placed at the entrance pupil of the telescope and consists of two
holes, 8 cm in diameter each, with centers separated by 19.2 cm. The light
through one of the holes is deflected by a wedge lens, so that a stellar 
wavefront produces two images in the focal plane of the telescope. The
two visible images (camera response peaks at 0.5 ~\micron) are recorded by 
the CCD camera (pixel size 10 ~\micron, corresponding to 0.74" at {\it f10}) 
in frames taken in succession at the maximum rate allowed by the camera 
electronics, i.e. about once per second. The exposure time of each frame 
alternates between 10 and 20 msec. As turbulent eddies cross the line of 
sight of the stellar source, the two images move differentially as the 
optical paths through the two apertures sample different
sections of the wavefront, with a separation which is commensurate with
$r_\circ$. On the other hand, tracking errors and wind buffeting of the
telescope assembly affect equally the motion of the two stellar images 
in the CCD field. The fluctuations in the differential motion of the two 
images are related to $r_\circ$ as follows.

Let $d$ be the separation between the two holes and $D$ the diameter of
each hole. Image motion due to turbulence arises from the corrugation in 
the wavefront, as it reaches the aperture mask: the normal
to the wavefront surface is the instantaneous angle of arrival. The 
covariance of the fluctuations of the angle of arrival can be related to 
the wavefront phase error fluctuations and, consequently, to the index of 
refraction structure parameter defined in the previous section, as discussed
by Sarazin \& Roddier (1990). The covariance in the fluctuations of the 
angle of arrival can be converted to a variance in the image motion, which 
along the direction connecting the two apertures (longitudinal) is shown by
Sarazin \& Roddier to be
\be
\sigma_l^2 = 2\lambda^2r_\circ^{-5/3}~[0.179 D^{-1/3} - 0.097 d^{-1/3}]
\label{eqsigmal}
\ee
while the variance in the orthogonal (transverse) direction is
\be
\sigma_t^2 = 2\lambda^2r_\circ^{-5/3}~[0.179 D^{-1/3} - 0.145 d^{-1/3}]
\label{eqsigmat}
\ee
Here, lengths are in cm and units of $\sigma$ are seconds of arc.
If the observations are made at a zenith angle $ZA$, the full width at half
maximum of the stellar image can be obtained independently from 
Equation ~\ref{eqsigmal} ($i=l$ below) and Equation ~\ref{eqsigmat} ($i=t$ 
below), as discussed in the mentioned source:
\be
\theta_{fwhm} \propto \lambda^{-1/5}~[\sigma_i^2 \cos (ZA)]^{3/5}
\label{eqsee}
\ee

The variance of the image motion is obtained from $N$ short exposures,
separately for the 10 msec and the 20 msec series (10 and 20 msec exposures 
are taken alternately). Since the statistical properties of the atmosphere 
do not change significantly over time scales on the order of 1--2 minutes, 
the $\theta_{fwhm}$ can be averaged over a fairly large number $N$ of 
exposures of each series. We thus can obtain seeing estimates derived from 
10 msec exposures, from 20 msec exposures (which typically yield lower 
values of $\theta_{fwhm}$ because the image motion has been smeared by the 
longer exposure) and extrapolations to ``zero exposure'', obtained by 
multiplying the 10 msec seeing by the ratio of the 10 and 20 msec measurements. 
This is based on the results of Martin (1987), where it is shown that for 
experimental configurations such as ours, the variation of the attenuation 
of the variance with exposure time is close to exponential. If we assume
that $\sigma^2(t) = \sigma^2(0) e^{-at}$, measurements at $t$ and $2t$ can
be combined to obtain $\sigma(0)=\sigma(t)[\sigma(t)/\sigma(2t)]$.
In the unusual instance in which the 10 msec $\theta_{fwhm}$ is smaller than 
the 20 msec $\theta_{fwhm}$, for the zero exposure $\theta_{fwhm}$ we take
the average of the two nonzero exposure measurements. In the next Section,
we will present statistics for the three exposure times separately, for
ease of comparison with monitors that measure image motion at a fixed
exposure time.

The technique assumes that the perturbed wavefront is swept across the aperture 
in its frozen form during the exposure (``Taylor's hypothesis''). Since
{\it differential} image motion is recorded by the DIMM, the instrument does
not sense turbulent eddies much larger than $d$, which produce the same
distortion over the two apertures. The differential image motion is thus 
determined by the flow, at the wind speed, of small--scale eddies, which
takes place on short time scales. If $w$ is the wind velocity at the perturbing 
layer, ideally the exposure time should then be $\ll d/w$; practical 
limitations are imposed by the clock resolution of the CCD and the
necessity to detect enough photons so that centroiding of the stellar
image is reliable with readily observable stars at any time. Note 
that an eddy of $r_\circ = 20$ cm ($\theta_{fwhm}=0.5''$ at
$\lambda =0.5$ \micron) will cross the line 
of sight in 20 msec, in a 10 m sec $^{-1}$ wind. Alternating exposures of 
10 and 20 msec, it is possible to obtain an indication on how far the 
atmosphere departs from the assumption mentioned above, and extrapolate 
to infer an approximation of the ideal case of ``zero 
exposure seeing''. The accuracy of the final seeing figure depends on $N$, 
$d$, $D$, $r_\circ$, the stability of the atmosphere and the precision of 
the instrumental setup, mainly the telescope focus which also determines the 
scale of the image in the focal plane. 

As mentioned above, differential image motion is insensitive to turbulent 
scales significantly larger than $d$. This may introduce a bias in carrying 
the seeing estimates inferred from a DIMM to those expected for a large 
telescope of diameter close to the upper limit of the Kolmogorov inertial range. 
It appears clear now that a finite outer scale $L_\circ$ of 
turbulence exists, varying principally between 10 and 25 m 
(Martin et al. 1999; Avila 2000; Bouzid 2000; Linfield et al. 2001), 
which is commensurate with the size of large telescopes. In fact, the 
seeing at the VLT is often significantly better than that 
obtained by a DIMM next to the telescope; that difference is 10\% in the 
optical and even more in the NIR (Sarazin, personal communication).

From Sarazin \& Roddier (1990), the error in the measurement of 
the image motion variance is given by 
\be
{\delta \sigma_i^2 \over \sigma_i^2} \simeq \sqrt {2\over N-1}
\ee
which can be converted to an error on $\theta_{fwhm}$ via equation \ref{eqsee}:
\be
{\delta \theta_{fwhm} \over \theta_{fwhm}} \simeq {3\over 5}\sqrt {2\over N-1}
\ee
In mild wind conditions and for 
$r_\circ < 40$ cm ($\theta_{fwhm} > 0.25"$) errors on $\theta_{fwhm}$  
smaller than 10\% can be obtained from the DIMM, by averaging $N\sim 100$
exposures. The
instrumental setup does however rapidly lose accuracy for $r_\circ > 40$.
Martin (1987) has investigated thoroughly the accuracy of seeing derived from 
the differential image motion technique. 

We average seeing measurements over intervals of up to 8 minutes in 
time ($N\sim 100$), providing a seeing figure to compare with those of
astronomical imaging exposures of typical duration. Independently, longitudinal and transverse image motion is monitored and seeing derived
from each; we verify that the two indications yield statistically the
same seeing and then averaged together. All seeing figures 
mentioned in this report refer to a wavelength of 0.5 ~\micron ~and are 
corrected to a zenithal line of sight. Both DIMM telescopes were mounted 
on a rigid platform which elevated the aperture to approximately 2.5 m
above the ground. Every time measurements were made, the DIMMs were in thermal 
equilibrium with the ambient air.

\subsection {Instrument Comparison}

\begin{figure}
\plotone{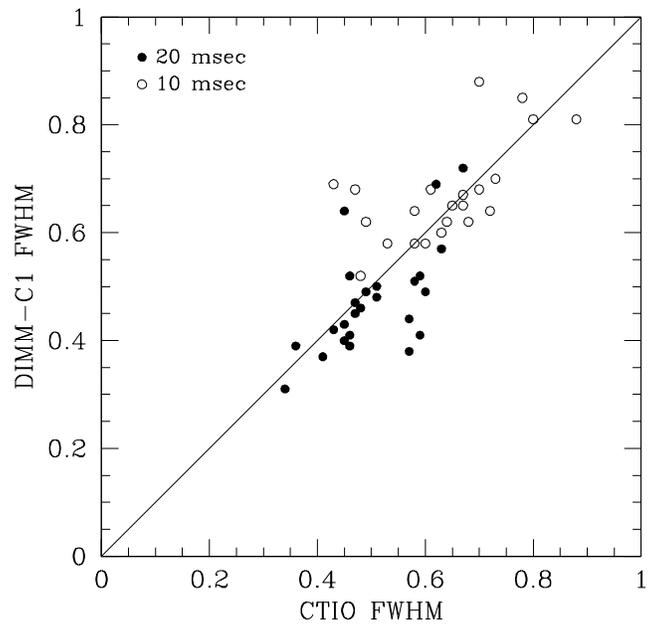}
\caption{Comparison of seeing measurements carried out with DIMM--C1
and the CTIO DIMM on two nights of October 2000. \label{dimm_comp}}
\end{figure}

Comparisons were made between DIMM--4 and DIMM--C1 for two nights. The
results of the two were identical to within the measurement errors, as should 
be expected because both the hardware and the software of the two devices are
identical. More important was the comparison carried out during two nights 
in October 2000 between DIMM--C1 and a DIMM brought to Atacama by a Cerro
Tololo Interamerican Observatory team (R. Blum, M. Boccas, E. Bustos and
B. Gregory). The CTIO DIMM had a higher data rate but it only integrated
at one exposure time, which was set to 20 msec on the first night and to
10 msec on the second. The seeing measurements could thus be compared 
between the two devices, which were different in both hardware and software.
Figure \ref{dimm_comp}~ illustrates the comparison of the two sets of data.
The two devices were placed to within 10 m of each other, at the summit
of Cerro Honar, at 5400 m. The aperture of
the CTIO DIMM was closer to the ground (approximately 1 m) than DIMM--C1
(approximately 2.5 m); both were within 5 m of the mountain edge, in the
direction of the incoming wind. A wind screen of negligible thermal capacity 
protected the CTIO DIMM; no screen protected DIMM--C1, which was more
vulnerable to wind shake; this produced numerous interruptions of data
taking and consequent reduction in the number of averaged records $N$ and
increased error in the DIMM--C1 measurements. The comparison between the 
two data sets reveals no noticeable bias and good agreement to within the
expected errors. The split between the 10 msec and 20 msec seeing
values is due in part to the fact that conditions were better during the
night of 20 msec comparison, and in part to the fact that 20 msec exposures
smear image motion and produce lower seeing values than those derived from
image motion obtained from shorter exposures, as discussed in the preceding 
Section.

\subsection {Measurement Strategy and Routines}

The ultimate purpose of our measurements is to establish the statistical 
properties of $\theta_{fwhm}$ at the best sites in the National Science 
Preserve region in the Atacama. 
A number of constraints are imposed on our campaign strategy. Given the
high altitude and budgetary restrictions, it is impractical to consider 
{\it extended} measurement campaigns on summits to which vehicular access is 
currently impossible. While the construction of a robotic DIMM is now under
way in Ithaca, obtaining a device that will work reliably and unattended at a 
remote site requires extensive testing and substantial resources. A robotic 
device will also be less portable, so its use should be aimed at testing 
the long term characteristics of a well chosen site.

Transportation of equipment and power sources restricts the ability to 
perform measurements to locations on the plateau and to a few elevations 
surrounding them. The latter include (a) two locations (one on Cerro Toco 
and one on Cerro Chajnantor) at which now abandoned sulphur mines were 
operated in previous decades, (b) the summit of the chain of Cerros de Honar, 
to which we have recently opened a rudimentary access road and (c) Cerro 
Chico, to which a small communications repeater was installed by a gas 
pipeline construction company. Access can be negotiated with rental 
four--wheel drive vehicles to those four locations. Those in Toco and 
Chajnantor, however, do not lead to prime locations for seeing measurements: 
neither reaches the summit and, in both cases, the track is on the east side,
and thus on the prevailing wind shadow of the mountain mass. In addition to 
strong disturbances in the air flow produced by that circumstance, 
katabatic winds are also a concern in those two cases. For Cerro Chico and
Cerros del Honar, however, vehicular access to the summits is possible. 
Both have a relatively clear immediate western horizon, although
Cerro Negro and Cerros de Mac\' on, almost as high as Chico and only 300 m
lower than Honar, are a few km away in the direction of prevalent incoming winds.
Moreover, Chico is located only 2 km from the foot of Cerro Chajnantor, 
and the effect of katabatic winds from Chajnantor, a significantly more 
substantial and higher mountain mass, is a concern.
Access to the other summits (Chajnantor, Toco, Negro and Chasc\' on) at this
time requires foot hikes several hours long. The construction of even a rudimentary
drivable track to any of those summits would be expensive, as would the 
transportation of equipment by means of a helicopter suited for high altitude 
flight. 

Given the aforementioned circumstances, a seeing campaign strategy was developed
that would follow three phases:  

\noindent I. We would first establish a reference frame for seeing statistics 
in the region at an easily accessible site, possibly above the local boundary 
layer. We would determine reliably the site's average seeing properties and 
seasonal variations, by carrying out a series of seeing runs spread over the 
seasonal cycle. This phase requires the deployment of a single DIMM (DIMM--4).

\noindent II. Next, we would compare 
characteristics of potentially attractive sites by means of relatively 
brief runs of seeing measurements at those sites, simultaneous with 
measurements at the reference site. This phase requires the deployment of
two DIMMs (DIMM--4 and DIMM--C1)

\noindent III. Finally, once the site with the best comparative characteristics 
is identified, a long--term campaign of continuous monitoring
with a robotic DIMM would be carried out. This phase requires the deployment
of a robotic DIMM, currently under construction.

Given its accessibility, possible partial emergence above the atmospheric 
boundary layer and central location in the Science Preserve Area, we chose 
Cerro Chico as the site at which a seeing reference standard for the region 
would be established (the road to Honar was not completed until August 2000). 
Observing consisted of runs of 
8--10 days duration each, spaced by approximately two months. Seven such runs
have been carried out, sufficient to fairly characterize the
seeing at Chico. This series of runs was preceded by a single run at the plateau 
level in May 1998, in the course of which DIMM--4 was deployed at the NRAO  
testing site (``ALMA container'', position of label 2 in Figure \ref{satview}),
at an elevation of 5050 m. This decision was made in order to ascertain our 
and the equipment's ability to effectively function at high altitude. Caution 
advised that such verification be made near a shelter. Successively, runs 
were carried out at the summit of Cerro Chico, some 2 km NW and about 100 m 
higher than the ALMA container. A list of the runs and their durations is 
given in Table 1. Columns 1 and 2 list the date of each run and the location; 
column 3 lists the number of nights ($N_n$) and the number of hours
($N_h$) in which useful data were collected; columns 4--12 list the 25\%, 50\%
and 75\% quartile values of $\theta_{fwhm}$, measured respectively for the
extrapolated ``zero exposure seeing'' (cols. 4--6), the 10 msec exposures 
(cols. 7--9) and 20 msec exposures (cols. 10--12). The coordinates of the two 
sites are given in Table 2. Atacama is in the Universal Transverse Mercator 
(UTM) zone nr. 19.

During operation, the DIMM aperture stands some 2.5 m above the ground. 
During the May 1998 run, the DIMM was placed about 5 m away from the shipping 
container that serves as shelter for NRAO's ALMA monitoring station. During 
observations, the DIMM was always upwind from the container. During the 
observations at Chico, the DIMM was placed on the summit of this modest 
orographic formation, at less than 5 m from its western edge, which slopes 
downhill rather steeply in the prevailing direction of incoming winds. This 
choice was motivated by the results of the simulations by De Young \& Charles (1995)  
of airflow over telescope sites, and the experience of ESO at Paranal and La 
Silla, which indicates that the effect of ground layer turbulence can be 
minimized by placing the DIMM near the edge of a summit, facing the incoming 
direction of prevailing winds. According to Martin et al. (2000), the median
contribution to the seeing of the ground boundary layer, between 2 m and 21 m 
above the ground, is 7\% of the total, with most of it occurring between 
2 and 7 m. For a median seeing near 0.7", the contribution of the ground
boundary layer alone would be on the order of 0.25", assuming that each layer's
contribution adds quadratically. Given our telescope setup, 
some measure of degradation of the seeing due to ground boundary layer should 
then be expected. In normal conditions, such deterioration is probably below the 
10\% level. During exceptionally good nights, however, the effect of the 
ground boundary layer may be more important. 

During measurements, we power the DIMM hardware off a regular automobile 
battery, which is charged daily by connecting it to the engine of a 
vehicle. Each evening after sunset, telescope setup takes approximately 30 
minutes. Each observing shift involves two persons. A first or second 
magnitude star is acquired, preferably to the south of zenith, so that the 
tracking errors resulting from misalignment between the telescope's and 
the Earth's polar axis are minimized. No more than 2 target stars per night 
are necessary; they are tracked at air masses generally between 1.1 and 1.6. 
Automatic guiding is provided by the DIMM 
software, and in routine operation, no attention is required, except for
the impact of exceptional environmental effects (wind gusts, frost forming
on the telescope mask, etc.) or occasional electronic glitches. DIMM data 
consists of a time series of (a) image motions, both longitudinal and 
transverse to the direction of the axis of the two holes in the mask; 
these are converted to seeing figures via Equation ~\ref{eqsee}; (b) a 
scintillation index, which is the variance of the flux associated with 
each image; (c) time, zenith angle, focus, image scale and ancillary
instrumental parameters. Roughly one record per second is acquired. 
Exposures of 10 msec and 20 msec are taken alternately, and separately 
analyzed. Off--line, the data is extrapolated to ``zero  exposure'', as 
described in the preceding section, and averaged over intervals of between 
2 and 8 minutes.

\subsection {Seeing Measurements: Results}

\begin{figure}[t]
\plotone{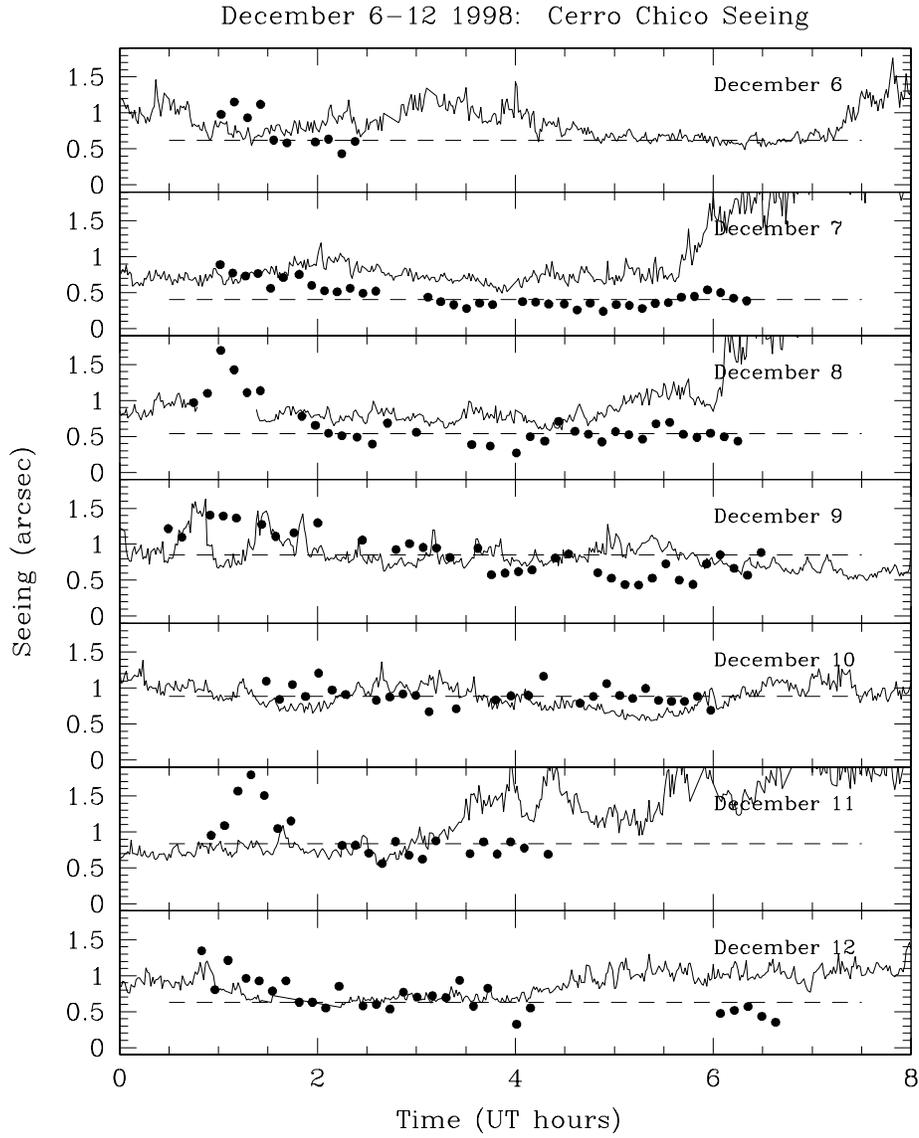}
\caption{Time series of seeing measurements for the December 1998 run. 
Filled circles are 8--minute seeing averages at Cerro Chico; thin solid
lines represent simultaneous seeing at Cerro Paranal. Dashed lines are
nightly Cerro Chico medians. Analogous plots for other runs can be
seen at {\it http://www.astro.cornell.edu/atacama/}.
Local midnight occurs at 4.31$^h$ UT. \label{see_dec}
}
\end{figure}

Figure \ref{see_dec} displays time series of ``zero exposure seeing'' data for 
each day of the December 1998 run. Similar plots for the other runs can be seen 
in {\it http://www.astro.cornell.edu/atacama}. Solid symbols identify 8 minute
seeing averages on Cerro Chico, while the horizontal dashed lines indicate
the daily median seeing. The thin solid line in each panel is 
a running mean of the Cerro Paranal seeing measurements obtained with a DIMM
similar to DIMM--4, at the same time as our data. Note that local midnight is 
at UT $=4.5^h$.

As illustrated in Table 1, the statistical properties of each of the Cerro
Chico runs are rather similar. The median values of the 0 ms seeing vary 
between 0.65" and 0.76", those of the 10 msec seeing between 0.56" and 0.65"
while those of the 20 msec seeing between 0.48" and 0.56". Quartiles of
0.56", 0.71" and 0.87" for the overall data set, including 38 nights and
153 hours of data, appeared to be closely mimicked by the corresponding
values for each run. In contrast, the May 1998 run at the plateau level
gave significantly higher quartile values of 0.93", 1.09" and 1.28".

\begin{figure}[t]
\plotone{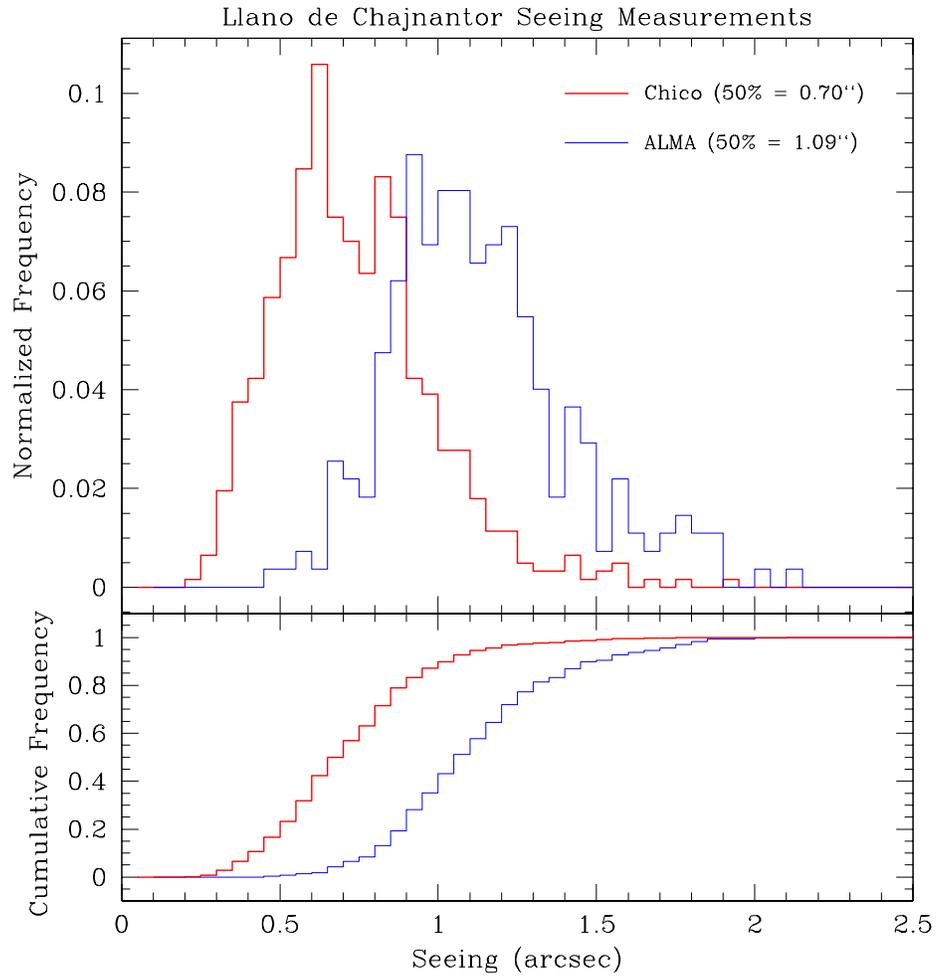}
\caption{Comparison of seeing  at the ALMA Container and
Cerro Chico, including all of 1998 Cerro Chico runs and May 1998 ALMA run. 
\label{see_alma_chico}}
\end{figure}

Figure \ref{see_alma_chico} displays a comparative summary of the seeing 
measurements at the ALMA container and at Cerro Chico. The distributions for 
the two sites are significantly different. It should be remarked that the 
Cerro Chico and ALMA container measurements are not simultaneous. 

The seeing at the beginning of the evening tends to be of lower quality. Such
measurements are generally made still in twilight conditions, as there are
advantages to doing the telescope setup in sunlight. The observation of
significant improvement occurring 1--2 hours later is also a common experience. 
This effect appears to be correlated with a decrease in the wind speed, which 
as we have seen usually occurs 1--2 hours after sunset. The second part of the 
night tends to have better seeing than the first; yet, the larger fraction 
of Cerro Chico data correspond to the first half of the night.

Some data exist of simultaneous seeing and
precipitable water vapor measurements. The latter are discussed in Paper II.
Eight reliable simultaneous measurements are currently available. Of those,
the four with total precipitable water vapor PWV$<1$ mm correspond to an 
average seeing of 0.51" at 0.5 \micron, while the other four, with PWV$>1$ 
mm, correspond to an average seeing of 0.84". For the first set of four, the
water vapor layer was tightly packed at low altitudes above the
ground, 50\% of PWV being in each case contained within 500 m from the 
plateau level. For the second set of measurements, those with high PWV, the
water vapor was distributed more widely in the vertical direction, the 50\%
boundary being above 1000m in each case. This correlation is suggestive but
statistically still quite weak. 

\begin{figure}[t]
\plotone{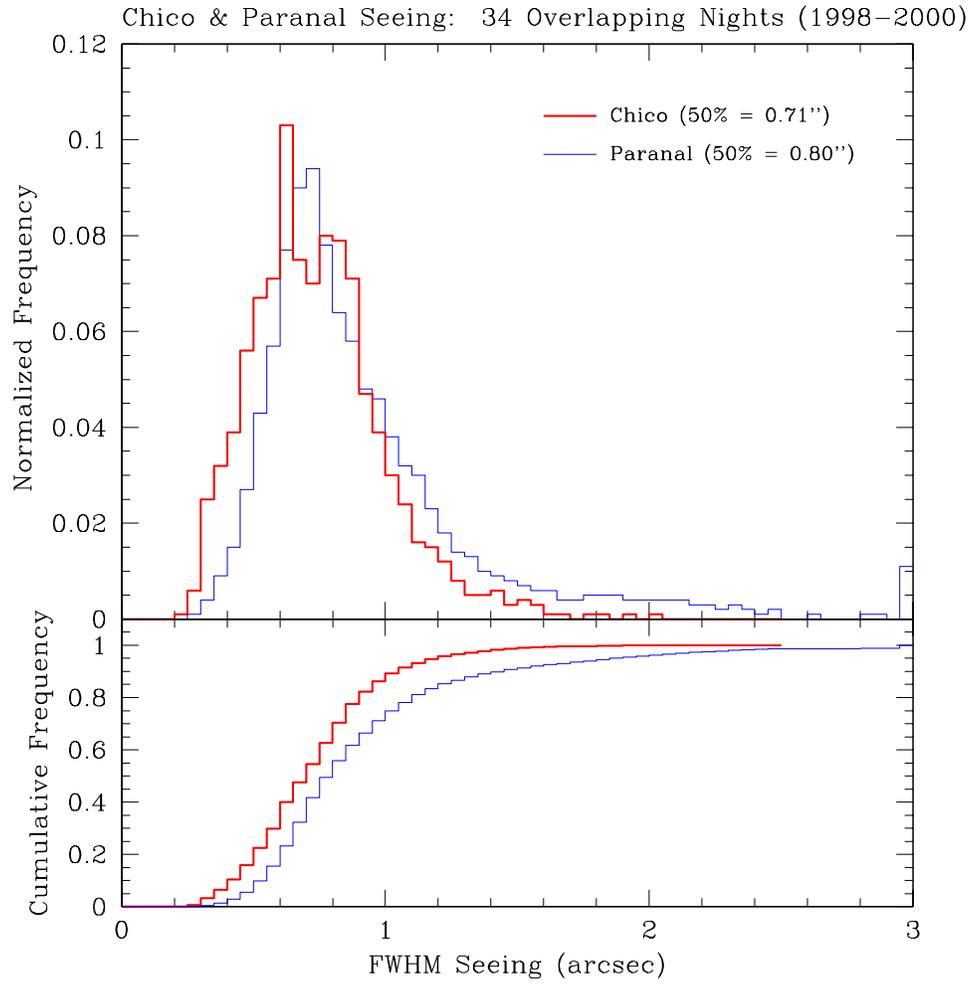}
\caption{Comparison of seeing  at the Cerro Paranal and
Cerro Chico. 
\label{see_chico_par}}
\end{figure}

While there should be no expectation for Cerro Chico to deliver the best 
seeing in the National Science Preserve region, it is useful to compare it
to other prime astronomical sites. The established astronomical 
site with the best historical seeing record, as measured with a similar
technique to our own is Cerro Paranal. Cerro Paranal boasts a ``zero exposure 
seeing'' median of $\theta_{fwhm}=0.66$", over several years of systematic 
DIMM tests (see {\it http://www.eso.org/gen-fac/pubs/astclim}). The historical 
median at La Silla, by comparison, is 0.87". The La Silla and Paranal measurements 
were obtained with a DIMM analogous to the one we used in Chajnantor and 
similarly calibrated. Cerro Paranal is located some 250 km to the west and 
180 km to the south of the Llano de Chajnantor. Figure \ref{see_chico_par} shows a comparison
of the seeing statistics for Cerro Chico and Cerro Paranal, for 34 simultaneous 
nights of observations. Cerro Chico's seeing ($\theta_{fwhm}=0.71$") is about 
12\% better than that at Paranal ($\theta_{fwhm}=0.80$"). During the 
period of these measurements, however, Cerro Paranal's seeing was about 21\% 
worse than its historical median. This circumstance has been noted by Sarazin 
\& Navarrete (1999), and ascribed to the exceptional {\it El Ni\~ no/La Ni\~ na} 
anomaly discussed in Section 3, which should have affected the Chico seeing 
as much a it affected Paranal's.
There has been an improving trend in the Paranal seeing in the last 
few months of 2000, bringing it to match its historical performance. Such an 
improvement was not noticeable at Chico, at least for the limited duration of 
the October 2000 run.

\section {Photometric Conditions}

We have not conducted a strict, quantitative survey of the photometric
conditions at Chajnantor. We have however kept a visual record of the sky
conditions, based on the substantial, combined experience of participants
at the seeing runs and anchored by the individual estimates of RG, 
who took part in all the runs. Of the 84 nights with good visual records, 
53 (63\%) were deemed photometric, 15 (18\%) were not photometric, with 
partial cloudiness, but suitable for other astronomical work, and 16 (19\%) 
were more than 60\% overcast. This included a full run of 11 cloudy nights 
(March 1999). 

This record of observations yields a lower fraction of clear nights than
obtained from historical satellite data records by Erasmus (2000), whose
results based on 6.7 \micron ~GOES 8 satellite imagery indicate that the 
atmospheric layers above 7500 m altitude in the Chajnantor region were
clear of clouds (cloud cover less than 70\% of sky) more than 90\% of the 
time between January 1995 and February 1996. The disparity between our 
visual record and the Erasmus' results may reflect the {\it El Ni\~ no/La 
Ni\~ na} anomaly of 1998, the contribution of low lying clouds to sky
cover, or poor statistics for
our visual records. At any rate, it is unlikely that the fraction of 
photometric nights at Chajnantor can match the extraordinary record of 
Cerro Paranal.

\section {Discussion and Future Work}

We have obtained a reliable reference frame for measurements of seeing in
the Chajnantor region. Provided by observations at Cerro Chico over 38 nights 
spread over the seasonal cycle, this reference frame is currently being
used to compare different sites in the vicinity of the Llano de Chajnantor.
While Cerro Chico does not qualify as a site expected to deliver the best 
observing conditions among astronomically attractive sites in the region, 
it has easy access, it produces results with repeatable statistics and is 
thus a good choice for benchmarking quality comparisons through simultaneous 
measurements at different sites. It also turns out to 
deliver high quality seeing, comparable with that of some of the best 
observatories on Earth. This allows for good expectations for 
the quality of other sites in the region. The next phase of our campaign
involves a series of runs of simultaneous seeing measurements at Chico and
other summits, in the course of which we will learn the differential qualities
of the various sites, and we will be able to tie those to the increasingly
more robust data sample for Cerro Chico itself. 

\begin{figure}
\plotone{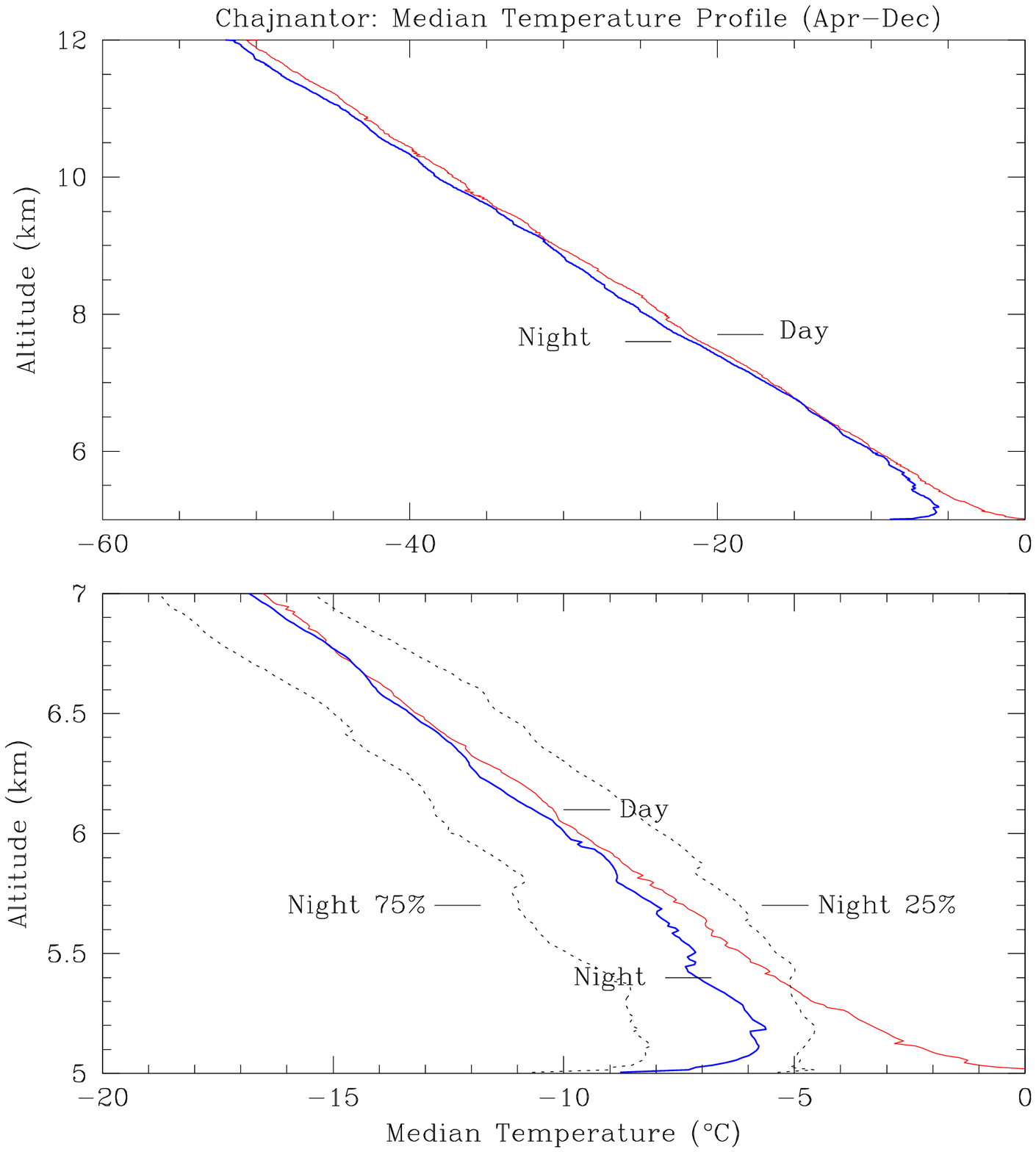}
\caption{Mean Temperature profile above Chajnantor. Night profile obtained
from 30 sonde flights; day profile obtained from 65 sonde fligths.
Dotted lines represent the 25\% and 75\% quartiles during night.
Sondes were launched between April and December. The lower panel shows
an expanded view of the lower 2 km of atmosphere.\label{mean_temp}}
\end{figure}

\begin{figure}
\plotone{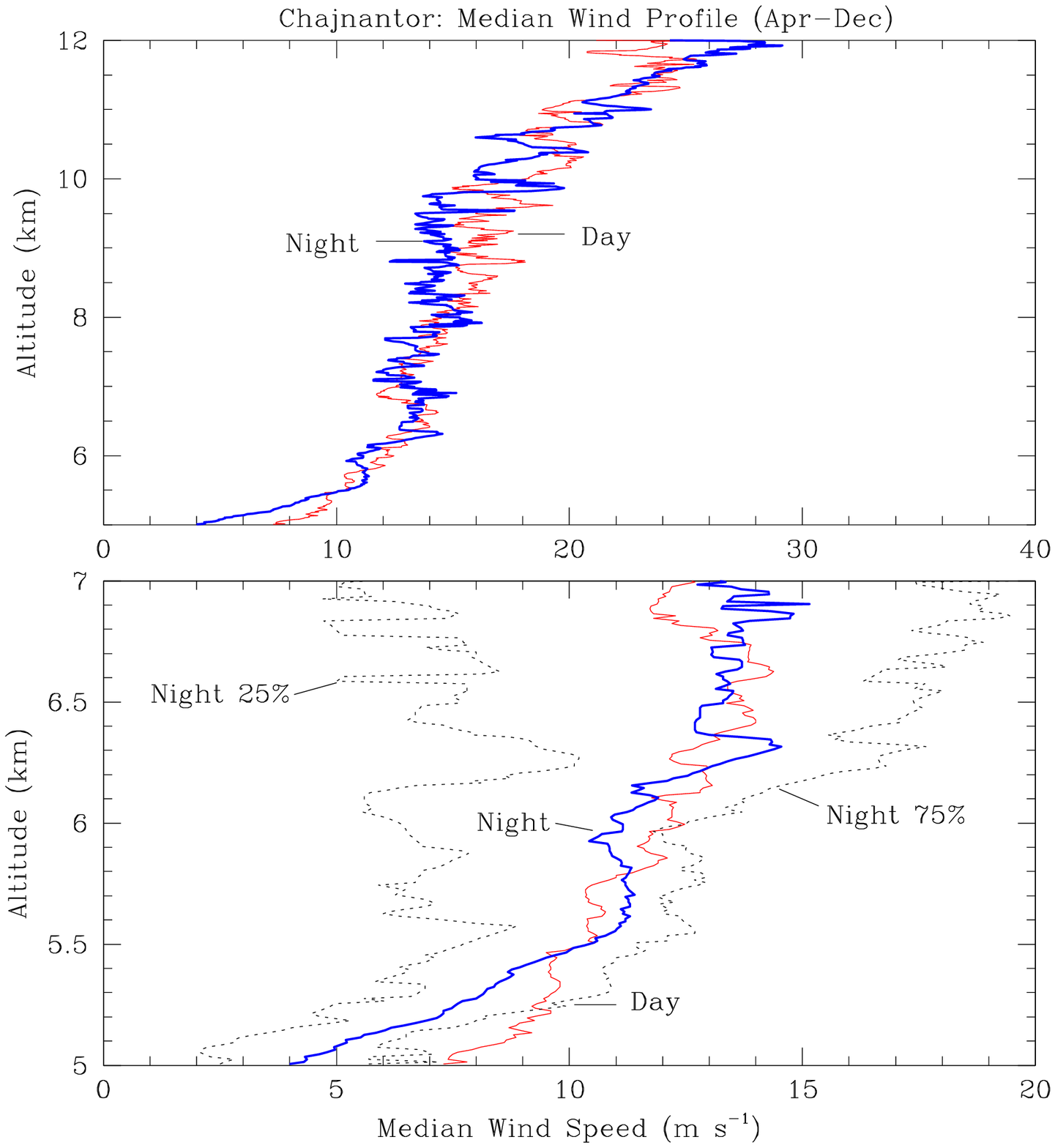}
\caption{Mean wind speed profile above Chajnantor. Night profile obtained
from 30 sonde flights; day profile obtained from 65 sonde fligths.
Dotted lines represent the 25\% and 75\% quartiles during night.
Sondes were launched between April and December. The lower panel shows
an expanded view of the lower 2 km of atmosphere. \label{mean_wind}}
\end{figure}

The difference between the seeing measurements at the ALMA Container and at 
Cerro Chico, a mere 100 m above the plateau level, suggests that 
boundary layer contributions dominate seeing at the plateau. This result was
partly expected, for air flow at the plateau level is strongly influenced by
thermally driven effects which, during daytime, result in strong winds from
the west and, during quiescent nights, in katabatic winds off the surrounding
mountain slopes. As expressed by the Richardson number, the onset of turbulence 
depends on the temperature gradient and the wind shear: the inhibition of 
convection produced by the positive thermal
gradient with increasing altitude normally occurring at night in the lower 
100 m of the atmosphere (Figure \ref{mean_temp}), is counterbalanced by strong 
wind shear (Figure \ref{mean_wind}). While the derivation of $C_n^2$ from
the coarse median profiles shown in Figures \ref{mean_temp} ~and \ref{mean_wind}~
is impossible due to their low resolution, it is well known (Hufnagel 1978) that 
the occurrence of temperature inversion layers in the troposphere tends to be 
accompanied with increased $C_n^2$ just below the inversion. While a good 
fraction of the boundary layer seeing clearly arises from the lowest 100 m 
over the plateau, as the comparison of the seeing measured at the ALMA Container 
and at Cerro Chico shows, an important fraction of it probably 
originates above the summit of Chico, for inversion layers between 200 and 
1000 m above the ground are common, as shown by radiosonde data in Paper II. Thus, optimistic expectations for high quality seeing at higher 
altitude peaks in the region are justified. 

The free atmosphere seeing arises mostly from temperature fluctuations 
associated with the tropopause and the jet stream; the median value of 
$\theta_{fwhm}$ has 
been estimated to vary between 0.31" at La Silla, 0.37" at the South Pole, 
and 0.40" to 0.46" at Mauna Kea, La Palma and Paranal. As for the boundary
layer, Marks \etal (1999) give estimates of its thickness at various 
astronomical sites. They range between 220 m at the South Pole, 900 m at 
La Silla and 2000 m at Cerro Paranal. At the Llano de Chajnantor region, 
the combination of ground surface layer and orographic boundary layer seeing 
may be quite complex: while the mean ground level rises gradually but steeply 
between the 2300 m of the Salar and the 5000 m of the plateau to the east, 
several hill formations stand between them, reaching elevations of about 5000 
m (Cerro Negro and Cerros de Mac\' on). Their influence on the thermal stability 
of the air flow is likely to be important, and dependent on the specific 
direction of the wind. A study of the characteristics of the air flow in 
the region by means of computer simulations is being carried out by D. De Young. 

At this stage, a few simple considerations can be made regarding 
expectations for potentially interesting sites in the region:
(i) it is likely that by accessing higher sites, the quality of the seeing
will improve, with respect to the already excellent values measured at Cerro
Chico; (ii) the infrared transparency improves with
elevation, as discussed in Paper II; (iii) such improvements 
will come with disadvantages, mainly that associated with progressively 
increasing high altitude discomfort and rapidly increasing 
wind speeds, which will place a burden on construction and operation of 
telescopes. The candidate sites for future testing include the summit of the 
Honar chain, Cerro Toco, Cerro Chajnantor, Cerro Chasc\' on and Cerro Negro. 
Measurements at Honar are already under way. 

The current, preliminary conclusions on the seeing conditions in the
Chajnantor region are as follows:

\begin{itemize}

\item The median seeing at 0.5 \micron ~over 7 nights, at the plateau level 
of 5050 m, is$\theta_{fwhm}= 1.1$". This is strongly affected by boundary layer contributions.

\item The optical seeing measured at Cerro Chico, 100 m above the plateau
level is $\theta_{fwhm}=$ 0.7" over 38 nights, between July 1998 and October 2000; this is 
12\% better than the seeing at Cerro Paranal. This result is likely to be an 
overestimate of the historical median for Cerro Chico, as in the same 
period the seeing at Paranal was 21\% higher than its historical median.

\item The percentage of optically photometric nights over the Chajnantor
region is near 60\%, while that of nights useful for astronomical observations
is close to 80\%.

\item Cerro Chico can now provide a benchmark against which to test the
quality of other sites in the region, via dual, simultaneous seeing 
measurements. Such efforts are under way.

\end{itemize}

{\bf Acknowledgments:} The support of Don Randel, Yervant Terzian, Joseph 
Veverka and Bryan Isacks of Cornell University, Riccardo Giacconi, Angel 
Otarola, Peter Shaver and Alvio Renzini of ESO, Martha Haynes of Cornell and 
Associated Universities, Inc., Robert L. Brown, Eduardo Hardy, Simon Radford 
and Geraldo Valladares of NRAO, don Tom\' as Poblete Alay and the staff of 
{\it La Casa de Don Tom\' as} are thankfully acknowledged. This study was made 
possible by a grant of the Provost's Office of Cornell University and a National Science Foundation  
grant AST--9910136.


\newpage

\begin{deluxetable}{cccrrrrrrrrr}
\tablewidth{0pt}
\tablenum{1}
\tablecaption{Seeing Data Summary \label{camps}}
\tablehead{
       & & & & 0 ms & & & 10 ms & & & 20 ms &  
\\[.2ex]
Date   & Location  & N$_n$/N$_h$ & $\theta_{25\%}$ & $\theta_{50\%}$ & $\theta_{75\%}$  & $\theta_{25\%}$ & $\theta_{50\%}$ & $\theta_{75\%}$ & 
$\theta_{25\%}$ & $\theta_{50\%}$ & $\theta_{75\%}$ 
}
\startdata
May98$^1$    & ALMA   & 5/41 & 0.93 & 1.09 & 1.28 & 0.85 & 0.97 & 1.12 & 0.72 & 
0.81 & 1.00 \nl
            &     &    &      &      &      &      &      &      &      &
     &      \nl
Jul98$^2$     & Chico  & 3/14 & 0.53 & 0.66 & 0.75 & 0.47 & 0.59 & 0.67 & 0.39 & 
0.49 & 0.59 \nl
Oct98     & Chico  & 7/33 & 0.52 & 0.70 & 0.86 & 0.46 & 0.59 & 0.73 & 0.39 &
0.48 & 0.64 \nl
Dec98     & Chico  & 7/31 & 0.50 & 0.66 & 0.90 & 0.44 & 0.56 & 0.75 & 0.38 &
0.48 & 0.66 \nl
Mar99$^3$     & Chico  &  1/2 &      & 0.70 &      &      &      &      &      &
     &      \nl
Apr00     & Chico  & 9/28 & 0.59 & 0.72 & 0.80 & 0.54 & 0.64 & 0.77 & 0.48 &
0.56 & 0.67 \nl
Jul00$^4$ & Chico  & 2/4  & 0.44 & 0.65 & 0.85 & 0.40 & 0.60 & 0.75 & 0.36 &
0.55 & 0.75 \nl
Oct00    & Chico  & 9/41 & 0.59 & 0.76 & 0.93 & 0.53 & 0.65 & 0.77 & 0.46 &
0.56 & 0.68 \nl
          &        &      &      &      &      &      &      &      &      &
          &        \nl
All       & Chico  &38/153& 0.55 & 0.71 & 0.87 & 0.49 & 0.61 & 0.75 & 0.42 &
0.52 & 0.66 \nl
\enddata
\tablenotetext{}  {N$_n$, N$_h$ are the total nr of nights, of hours observed.}
\tablenotetext{1} {Three nights lost to bad weather.}
\tablenotetext{2} {Data taking limited by equipment malfunction.}
\tablenotetext{3} {Bad weather for full run duration: Bolivian Winter conditions.}
\tablenotetext{4} {Data loss due to bad weather; data of dubious quality taken 3 other nights.}
\end{deluxetable}

\begin{deluxetable}{lccccc}
\tablewidth{0pt}
\tablenum{2}
\tablecaption{Site Coordinates \label{sites}}
\tablehead{
\colhead{Site}   & \colhead{Latitude}  & \colhead{Longitude}  & \colhead{UTM Northing}  
& \colhead{UTM Easting}   & \colhead{Elevation}   
}
\startdata
ALMA Container & S 23$^\circ$ 01'.2 & W 67$^\circ$ 45'.2 & 7453.7 km & 627.8 km & 5050 m \nl
Cerro Chico    & S 23$^\circ$ 00'.3 & W 67$^\circ$ 46'.3 & 7455.7 km & 626.2 km & 5150 m \nl
\enddata
\end{deluxetable}

\newpage

\newpage

\end{document}